\documentclass[a4paper,11pt]{article}
\pdfoutput=1 

\usepackage{jcappub} 

\usepackage[T1]{fontenc} 

\def\aj{{AJ}}                   
\def\apj{{ApJ}}                 
\def\apjl{{ApJ}}                
\def\apjs{{ApJS}}               
\def\aap{{A\&A}}                

\def\jcap{{J. Cosmology Astropart. Phys.}}
\def\mnras{{MNRAS}}             
\def\prd{{Phys.~Rev.~D}}        
\def\memsai{{Mem.~Soc.~Astron.~Italiana}}

\title{Revisiting the axion bounds from the Galactic white dwarf luminosity
  function}
\author[a,b,1,2]{M. M. Miller Bertolami\note{Corresponding author.}\note{On
    leave of absence from CONICET.}}
\author[b,c]{B. E. Melendez}
\author[b,c]{L. G. Althaus}
\author[d,e]{J. Isern}

\affiliation[a]{Max-Planck-Institut f\"ur Astrophysik,\\ Karl-Schwarzschild-Str. 1, 85748, Garching, Germany}
\affiliation[b]{Instituto de Astrof\'isica de La Plata, UNLP-CONICET,\\ Paseo del Bosque s/n, 1900 La Plata, Argentina}
\affiliation[c]{Grupo de evoluci\'on estelar y pulsaciones,\\ Facultad de Ciencias Astron\'omicas y Geof\'isicas, Universidad Nacional de La Plata,\\ Paseo del Bosque s/n, 1900 La Plata, Argentina}
\affiliation[d]{Institut de Ci\'encies de l'Espai (CSIC),\\ Facultat de  Ci\'encies, Campus UAB, Torre C5-parell, 08193 Bellaterra, Spain}
\affiliation[e]{Institute for Space Studies of Catalonia,\\ c/Gran Capit\'a
 2-4,  Edif. Nexus 104, 08034 Barcelona, Spain.}

\emailAdd{marcelo@MPA-Garching.MPG.DE}
\emailAdd{brenmele@gmail.com}
\emailAdd{althaus@fcaglp.fcaglp.unlp.edu.ar}
\emailAdd{isern@ieec.cat}

\abstract{It has been shown that the shape of the luminosity function of white
  dwarfs (WDLF) is a powerful tool to check for the possible existence of
  DFSZ-axions, a proposed but not yet detected type of weakly interacting
  particles. With the aim of deriving new constraints on the axion mass, we
  compute in this paper new theoretical WDLFs on the basis of WD evolving
  models that incorporate for the feedback of axions on the thermal structure
  of the white dwarf.  We find that the impact of the axion emission into the
  neutrino emission can not be neglected at high luminosities ($M_{\rm
    Bol}\lesssim 8$) and that the axion emission needs to be incorporated
  self-consistently into the evolution of the white dwarfs when dealing with
  axion masses larger than $m_a\cos^2\beta\gtrsim 5$ meV (i.e. axion-electron
  coupling constant $g_{ae}\gtrsim 1.4\times 10^{-13}$).  We went beyond
  previous works by including 5 different derivations of the WDLF in our
  analysis. Then we have performed $\chi^2$-tests to have a quantitative
  measure of the assessment between the theoretical WDLFs ---computed under
  the assumptions of different axion masses and normalization methods--- and
  the observed WDLFs of the Galactic disk. While all the WDLF studied in this
  work disfavour axion masses in the range suggested by asteroseismology
  ($m_a\cos^2\beta\gtrsim 10$ meV; $g_{ae}\gtrsim 2.8\times 10^{-13}$) lower
  axion masses can not be discarded from our current knowledge of the WDLF of
  the Galactic Disk. A larger set of completely independent derivations of the
  WDLF of the galactic disk as well as a detailed study of the uncertainties
  of the theoretical WDLFs is needed before quantitative constraints on the
  axion-electron coupling constant can be made.}

\begin{document}
\maketitle
\flushbottom

\section{Introduction}

Despite its success, the standard model of particle physics has some
unsolved issues. Among them, is the CP-problem of Quantum
Chromodynamics (QCD), i.e. the absence of a CP-violation in the strong
interactions. One elegant solution to the CP-problem is the
Peccei-Quinn mechanism \citep{1977PhRvL..38.1440P} in which the
coefficient of the CP-violating term of the QCD lagrangian is proposed
to be a dynamical field (the {\it axion field}) whose vacuum
expectation value naturally leads to CP-conservation. One of the
natural consequences of such mechanism is the existence of a new
particle, the {\it axion}
\citep{1978PhRvL..40..279W,1978PhRvL..40..223W}.  Soon after its
proposal it was realized that stellar astrophysics was an excellent
tool to constrain axion properties \citep{1978JETPL..27..502V}. While
the original axion models were soon ruled out by observations,
``invisible'' axion models such as the DFSZ \citep{Zhitnitsky,
  1981PhLB..104..199D} and KSVZ models
\citep{1979PhRvL..43..103K,1980NuPhB.166..493S} are much more elusive
(see \cite{1996slfp.book.....R} for a review). In particular DFSZ-type
axions couple to electrons and would be emitted from the interior of
white dwarfs (WD) and red giant cores, opening the possibility of using
those stellar populations as laboratories to constrain axion
properties (e.g. \cite{1986PhLB..166..402R,1992ApJ...392L..23I,2008ApJ...682L.109I}, and more
recently \cite{2013arXiv1308.4627V}). The
coupling strength between electrons and DFSZ-axions is defined by the
axion-electron adimensional coupling constant, $g_{ae}$, which is related to
the mass of the axion ($m_a$) through 
\begin{equation}
g_{ae}=2.8\times 10^{-14}\times m_{a[{\rm meV}]}\times \cos^2\beta
\end{equation}
where $\cos^2\beta$ is a model dependent parameter. $g_{ae}$ is also related
to the axionic fine structure constant $\alpha_{26}$ by
\begin{equation}
\alpha_{26}=10^{26}\times {g_{ae}}^2/4\pi .
\end{equation}

Because the evolution of white dwarfs is mostly a simple cooling process and
the basic physical ingredients needed to predict their evolution are
relatively well known, white dwarfs offer a unique opportunity to test new
physics under conditions that can not be obtained in present day laboratories
\cite{2004ApJ...602L.109W, 2011A&A...527A..72A,2013PhRvD..88d3517D}.  In the
last decade, the white dwarf luminosity function (WDLF) has been noticeably
improved by large sky surveys \citep{2006AJ....131..571H, 2008AJ....135....1D,
  2009A&A...508..339K, 2011MNRAS.417...93R}, leading to the possibility of new
studies of the impact of the axion emission in the WDLF. Based on these
improvements of the WDLF \cite{2008MmSAI..79..545I, 2008ApJ...682L.109I}
computed the impact of the axion emission in the WDLF by adopting a
perturbative approach on the axion emission.
\cite{2008ApJ...682L.109I,2009JPhCS.172a2005I} find that the inclusion of the
DFSZ-axion emissivity, with $m_a\cos^2\beta\sim 5$ meV, in the evolutionary
models of white dwarfs would improve the agreement between the theoretical and
observational WDLFs, and that axion masses $m_a\cos^2\beta> 10$ meV are
clearly excluded. Interestingly enough, these values were in concordance with
the values obtained from the secular drift of the period of pulsation of ZZ
Ceti stars
\cite{1992ApJ...392L..23I,2010A&A...512A..86I,2008ApJ...675.1512B}. However,
the recent analysis of the observed pulsation period of G117-B15A and R548
have provided a value of $m_a\cos^2\beta=17.4^{+2.3}_{-2.7}$ meV and
$m_a\cos^2\beta=17.4^{+4.3}_{-5.8}$ meV respectively
\citep{2012MNRAS.424.2792C, 2012JCAP...12..010C}. Interestingly enough, based
on an older determination of the observed period drift
\cite{2005ApJ...634.1311K} and simplified chemical profiles, a similar result
($10.4$ meV$\lesssim m_a\cos^2\beta \lesssim 26.5$ meV) was suggested by the
thick envelope solutions of \cite{2008ApJ...675.1512B}.  It should be noted
that for axion masses as high as $m_a\cos^2\beta\sim 17$ meV the axion
emission becomes the dominant cooling process down to very low luminosities
($M_{\rm Bol}\sim13$). In such a situation, we expect the existence of a
significant axion emission to impact the thermal structure of the white
dwarf. Thus, for the range of interest of the axion masses suggested by
asteroseismology the axion emission can not be treated as a perturbation to
the white dwarf cooling and a self consistent treatment of the axion emission
is necessary.  In view of the previous arguments, the inconsistency between
the masses derived by both methods calls for a reanalysis of the previous
results in the light of a self consistent treatment of the axion emission. In
addition, new determinations of the WDLF that extend to the high luminosity
regime have become recently available \citep{2009A&A...508..339K,
  2011MNRAS.417...93R}. At these luminosities the axion emission can be very
important even for low axion masses.  Hence, to make use of the new available
data, full evolutionary models derived from the progenitor history are also
needed.

In the present work, we analyze constraints on the axion mass by means of a
detailed analysis of the WDLF of the Galactic disk. We go beyond previous
works by studying the impact of the axion emission in the cooling of white
dwarfs by means of a self consistent treatment of the axion emission and state
of the art white dwarf models. In addition, we extend the scope of our work by
taking into account different derivations of the Galactic WDLF. In particular,
we include in the analysis an estimation of the WDLF of bright white
dwarfs. Also, we analyze to which extent the normalization procedure adopted
for the white dwarf luminosity function affect the results. The paper is
organized as follows: in the next section we describe the numerical tools,
input physics and initial white dwarf models. We also describe briefly the
method adopted to compute the theoretical white dwarf luminosity functions. In
Section \ref{impact}, we present the results of the impact of the axion
emission in the white dwarf models, paying special attention to the feedback
of the axion emission in the thermal structure of the white dwarf.  Next, in
Section \ref{axionWDLF} we quantitatively assess the impact of the axion
emission in the theoretical WDLF and compare with different empirical WDLFs.
Finally, in Section \ref{conclusion} we summarize our results and conclusions
and propose some future work that needs to be done in order to improve these
constrains.

\section{Numerical tools}
\subsection{Input physics and initial white dwarf models}
The stellar evolution computations presented in this work have been performed
with {\tt LPCODE} stellar evolution code, which has been used to study
different problems related to the formation and evolution of white dwarfs
---e.g. \cite{2005A&A...435..631A,2010ApJ...717..183R, 2013ApJ...775L..22M}. A
detailed description of the code is available in \cite{2010ApJ...717..183R}
and references therein, here we only mention those which are most relevant for
the present work. The main physical ingredients included in the simulations of
white dwarfs computed with {\tt LPCODE} comprise the following. The equation
of state for the high density regime is that of \cite{1994ApJ...434..641S}
while for the low density regime we use an updated version of the equation of
state of \cite{1979A&A....72..134M} (Mazzitelli 1993, private communication).
Conductive opacities are those of \cite{2007ApJ...661.1094C} while radiative
opacities are those of the OPAL project \citep{1996ApJ...464..943I}
complemented at low temperatures by the molecular opacities produced by
\cite{2005ApJ...623..585F}. White dwarf models computed with {\tt LPCODE} also
include detailed non-gray model atmospheres to provide accurate boundary
conditions for our models which include non-ideal effects in the gas equation
of state, see \cite{2012A&A...546A.119R} for details. Neutrino cooling by
Bremsstrahlung, photo and pair production are included following the recipes
of \cite{1996ApJS..102..411I}, while plasma processes are included according
to \cite{1994ApJ...425..222H}. All relevant energy sources are taken into
account in the simulations, including marginal nuclear burning, the release of
latent heat and the gravitational energy associated with the phase separation
in the carbon-oxygen profile induced by crystallization. The inclusion of all
these energy sources is done self-consistently and locally coupled to the full
set of equations of stellar evolution. In addition the effects of time
dependent element diffusion during the white dwarf evolution is also taken
into account following treatment of \cite{1969fecg.book.....B} for
multicomponent gases. It is worth noting for the aim of the present work
that {\tt LPCODE} has recently been tested against other
well-known stellar evolution code and it was found that uncertainties in white
dwarf cooling times arising from different numerical implementations of the
stellar evolution equations were below 2\% \citep{2013Salaris}. This, together with the state of
the art input physics adopted in {\tt LPCODE} provides a very solid ground for
the present work.

The initial white dwarf models adopted in our simulations were taken from
\cite{2012RomeroPhD} and \cite{2010ApJ...717..183R}. Specifically 4 different
initial white dwarf models of 0.524$M_\odot$, 0.609$M_\odot$, 0.705$M_\odot$
and 0.877 $M_\odot$ were adopted. These models were obtained from computing
the complete evolution of initially 1$M_\odot$, 2$M_\odot$, 3$M_\odot$ and 5
$M_\odot$ ZAMS stars with Z=0.01, which is in agreement with semi-empirical
determinations of the initial-final mass relationship
\citep{2009ApJ...692.1013S}. Note that physically sounding initial WD models
are relevant at relatively high luminosities. In fact, \cite{2013Salaris}
found that, at $\log L/L_\odot\gtrsim -1.5$, differences up to 8\% can be
found due to gravothermal differences in the initial white dwarf models, even
when the same chemical stratification is adopted.

\subsection{DFSZ-axion emission}

  For the sake of completeness DFSZ-type axion emission by both Compton
  and Bremsstrahlung processes were included in {\tt LPCODE}, although
  only Bremsstrahlung processes are relevant in white dwarfs. Also,
  expressions concerning the axion emissivity were included for both
  the strong   
($\Gamma > 1$)\footnote{Where the ionic coupling constant $\Gamma$ for multicomponent plasmas is defined following \cite{1994ApJ...434..641S} as 
$$
\Gamma=2.275\times 10^5 \frac{(\rho Y_e)^{1/3}}{T}
\sum_i (n_i/n_{\rm ions}) {Z_i}^{5/3},
$$
where the sum is taken over all considered nuclear species. }
   and weak ($\Gamma < 1$) ion-correlations and for both the strongly degenerate
  and non-degenerate regimes. In what follows, for the sake of
  clarity, we describe the adopted expression for DFSZ-axion
  emissivity. Axion Bremsstrahlung emission under degenerate
  conditions ($\epsilon_{\rm BD}$) was included adopting the prescriptions of
  \cite{1987ApJ...322..291N,1988ApJ...326..241N} for strongly coupled
  plasma regime ($\Gamma >1$)

 and \cite{1995PhRvD..51.1495R} for weakly
 coupled plasmas ($\Gamma <1$). Specifically, we computed 
\begin{equation}
\epsilon_{\rm BD}=10.85\ \alpha_{26}\ {T_8}^4 
\sum_j^{N_{\rm isot}} \frac{X_j {Z_j}^2}{A_j}\times F_j 
\end{equation}
were $F_j$ is given by \citep{1987ApJ...322..291N,1988ApJ...326..241N} for $\Gamma >1$ and in the case of $\Gamma <1$ is given by $F_j=F$ where  
\begin{eqnarray}
F&=&\frac{2}{3} {\rm ln}\left(\frac{2+\kappa^2}{\kappa^2}\right) \nonumber\\
&+&\left[\frac{2+5\kappa^2}{15}{\rm ln}\left(\frac{2+\kappa^2}{\kappa^2}\right)
-\frac{2}{3}  \right]\times {\beta_F}^2
\end{eqnarray}
with   $\beta_F$ and $\kappa$ given as
\begin{eqnarray}
\kappa^2&=& \frac{2\pi\alpha\hbar^3 c}{m_u k}\frac{\rho}{T} 
\sum_j \frac{X_j {Z_j}^2}{A_j} \frac{1}{{p_F}^2},\\ 
{\beta_F}^2&=&\frac{{p_F}^2}{ {m_e}^2 c^2+{p_F}^2}, 
{p_F}^2=\hbar\left(\frac{3\pi^2\rho}{\mu_e m_u}\right)^{1/3}.
\end{eqnarray}
For the non-degenerate regime ($\epsilon_{\rm BND}$),
 Bremsstrahlung was derived from the expressions presented in
 \cite{1996slfp.book.....R}, specifically we adopted
\begin{eqnarray}
\epsilon_{\rm B ND} &=& 5.924\times 10^{-4}\ \alpha_{26}
\frac{{T_8}^{5/2}\rho}{\mu_e}
\sum_{j=1}^{N_{\rm isot}}
\left[\frac{X_j}{A_j}\right] \nonumber
\\ 
&\times&
\left[{Z_j}^2\left(1-\frac{5}{8}\frac{\kappa^2\hbar^2}{m_e T k}\right)\right. \nonumber\\
&+&\left.  \frac{Z_j}{\sqrt{2}}\left(1-\frac{5}{4}\frac{\kappa^2\hbar^2}{m_e T k}\right)
\right]\ \ \ \left[\frac{\rm erg}{\rm g\ s}\right].
\end{eqnarray}
with the screening scale $\kappa$ computed as
\begin{equation}
\kappa^2=\frac{4\pi\alpha\hbar c}{kT} \hat{n}, \ \   
\hat{n}\simeq n_e +\sum_j^{N_{\rm isot}} {Z_j}^2 n_j.
\end{equation}

 Axion emission by Compton processes was
 included following \cite{1995PhRvD..51.1495R} taking into
 account the effects of Pauli blocking under degenerate conditions and relativistic corrections,
\begin{equation}
\epsilon_{\rm compton}= 33\ \alpha_{26}\ Y_e\  {T_8}^6\ F_c\ \ \ \ 
 \hbox{erg g$^{-1}$ s$^{-1}$}.
\end{equation}
Where, following  \cite{1995PhRvD..51.1495R}, we computed $F_c$ as
\begin{equation}
F_c=\left(1+ {F_{\rm comp. deg.}}^{-2}\right)^{-1/2}.
\end{equation}
with 
\begin{equation}
F_{\rm comp. deg.}=4.96\times  10^{-6} {\mu_e}^{2/3} \frac{T}{\rho^{2/3}}.
\end{equation}
Then, the total axion emission ($\epsilon_{\rm axion}$) was then computed as:
\begin{equation}
\epsilon_{\rm axion}=\epsilon_{\rm Brem}+\epsilon_{\rm compton}
\end{equation}
where the Bremsstrahlung emission was obtained from
\begin{equation}
\epsilon_{\rm Brem}=
\left(\frac{1}{\epsilon_{\rm B ND}}+
\frac{1}{\epsilon_{\rm B D}}\right)^{-1}
\end{equation}
$\epsilon_{\rm B D}$ was then computed as $\epsilon_{\rm BD\, (\Gamma<1)}$ if
${\Gamma}<0.9$, $\epsilon_{\rm BD\, (\Gamma>1)}$ if ${\Gamma}>1.1$ and
interpolated linearly between both expressions when $\Gamma \in (0.9,1.1)$. It
is worth noting that, while the details of the interpolation scheme will
significantly affect the axion emission in intermediate regimes the mass of
those regions is small and will not introduce any significant change in the
global cooling speed of the white dwarf.

\subsection{Theoretical white dwarf luminosity functions}
As mentioned in the introduction, with the current knowledge of the WDLF it is
possible to put constraints on the axion mass independently from those
coming from asteroseismoogy. In particular, as the WDLF reflects the
global properties of the whole population of white dwarfs we expect it to be
less sensitive to the accuracy of our understanding of the internal structure of
individual white dwarf stars, as it is the case in asteroseismological
determinations of the axion mass. Of course, this advantage is obtained at the
price of relying on our present knowledge of stellar population properties
such as initial mass functions (IMF) and galactic stellar formation rates
(SFR) which do not play any role in asteroseismological
determinations. Fortunately, as shown by \cite{2008ApJ...682L.109I}, the
details of those ingredients do not seem to play an important role if the
range of WD luminosities used for WDLF comparisons is appropriately chosen.

To construct theoretical white dwarf luminosity functions we adopted
the method presented by \cite{1989ApJ...341..312I}. Thus, the number
of white dwarfs per logarithmic luminosity and volume is computed as
\begin{equation}
\frac{dn}{dl}=-\int_{M_1}^{M_2} \psi(t) \left(\frac{dN}{dM}\right)
 \left(\frac{\partial t_c}{\partial l}\right)_m dM
\label{eq:dndl}
\end{equation}
where $\psi(t)$ is the galactic stellar formation rate at time $t$,
$N(M)$ is the initial mass function and $t_c(l,m)$ is the time since
the formation of a white dwarf, of mass $m$, for the star to reach a
luminosity $log(L/L_\odot)=l$. In order to compute the integral in
equation \ref{eq:dndl} we also need the initial-final mass relation
$m(M)$, and the pre-white dwarf stellar lifetime $t_{ev}(M)$. It is
worth noting that, for a given white dwarf luminosity ($l$) and mass
of the progenitor ($M$) the formation time of the star, $t$, is
obtained by solving
\begin{equation}
t+t_{ev}(M)+t_c(l,m)=T_{OS},
\label{eq:time}
\end{equation}
where $T_{OS}$ is the assumed age of the oldest star in the computed
population. The lowest initial mass that produces a white dwarf with
luminosity $l$ at the present time ($M_1$) is obtained from eq. \ref{eq:time}
when $t=0$. The value of $M_2$ corresponds to the largest stellar mass
progenitor that produces a white dwarf.
In order to compute the impact of axions on the white dwarf luminosity
function, for each initial white dwarf model, 8 cooling sequences with
different assumed axion masses were computed ($m_a$
 $\cos^2\beta$= 0, 2.5, 5, 7.5,
10, 15, 20 \& 30 meV). In addition, to compute eq.  \ref{eq:dndl} we
adopt a Salpeter initial mass function, the initial-final mass
relation from \cite{2009ApJ...692.1013S}, the stellar lifetimes from
the BaSTI database \cite{2004ApJ...612..168P} and constant star
formation rate.

As shown by \cite{2008ApJ...682L.109I} the bright part of the WDLF is almost
independent of the stellar formation rate or the age of the disk, as their
main effects are absorbed in the normalization procedure. For the sake of
comparison with \cite{2008ApJ...682L.109I} for most of the computations we
compared the theoretical and observationally derived WDLF within the range
$-1\leq log(L/L_\odot)\leq -3$ ($7.25 \leq M_{\rm Bol}\leq
12.25$)\footnote{Throughout this work the assumed relationship between the
  bolometric magnitude and the luminosity of the star is adopted consistently
  with the data of \cite{2006AJ....131..571H} (i.e. $M_{\rm Bol}=-2.5
  \log(L/L_\odot)+4.75$).}.  A disk age of 11 Gyr is assumed throughout the
present work ($T_{OS}=11$ Gyr). Regarding the normalization of the theoretical
WDLF we tried two different approaches. First we followed
\cite{2008ApJ...682L.109I} and normalized the WDLF so that, for a given
$l_{\rm bin}$ value we have
\begin{equation}
\frac{dn}{dl}(l_{\rm bin})= n^{\rm O}(l_{\rm bin})
\label{eq:norm1}
\end{equation}
where $n^{\rm O}(l_{\rm bin})$ stands for the number of stars per volume, per
luminosity bin, inferred in Galactic WDLFs \citep{2006AJ....131..571H,
  2008AJ....135....1D, 2008ApJ...682L.109I} in the luminosity bin with
$log(L/L_\odot)=l_{\rm bin}$. This was done taking $l_{\rm bin}$ equal to all
luminosity bins presented between $l\simeq -2$ and $l\simeq -3$ (the exact
value depending on the binning of each WDLF). Only normalization points in the
low luminosity range should be preferred. Otherwise, axions are a dominant
cooling mechanism at the normalization luminosity and the differential effect
of the axion emission is hidden.  As will be seen in section \ref{axionWDLF},
the choice of the specific normalization point does not affect the main
conclusion of the work, but it has some impact in the quantitative comparison
between theoretical and observationally derived WDLF. In order to make the
theoretical WDLF less dependent on the particular normalization point we also
adopted a second normalization scheme. Specifically, we normalized the WDLF by
requiring the total number of stars per volume in a given magnitude range
($M_{\rm Bol}^1,M_{\rm Bol}^2$) to fit the observations , i.e.
\begin{equation}
\sum_{M_{\rm Bol}^i\in (M_{\rm Bol}^1,M_{\rm Bol}^2)} n^{\rm O}(M_{\rm
  Bol}^i) \Delta M_{\rm Bol}= \int_{M_{\rm Bol}^1}^{M_{\rm Bol}^2}
\frac{dn}{dl} dM_{\rm Bol}.
\label{eq:norm2}
\end{equation}
Our preferred choice is $M_{\rm Bol}^1=9.5\ (9.6)$ and $M_{\rm
  Bol}^2=12.5\ (12.4)$ for \cite{2006AJ....131..571H},
\cite{2008ApJ...682L.109I}, \cite{2011MNRAS.417...93R} and
\cite{2014A&A...562A.123M} \citep{2008AJ....135....1D} WDLFs.

\section{Impact of axion emission in white dwarf cooling}
\label{impact}
\begin{figure*}[ht]
\includegraphics[clip, angle=0, width=15.cm]{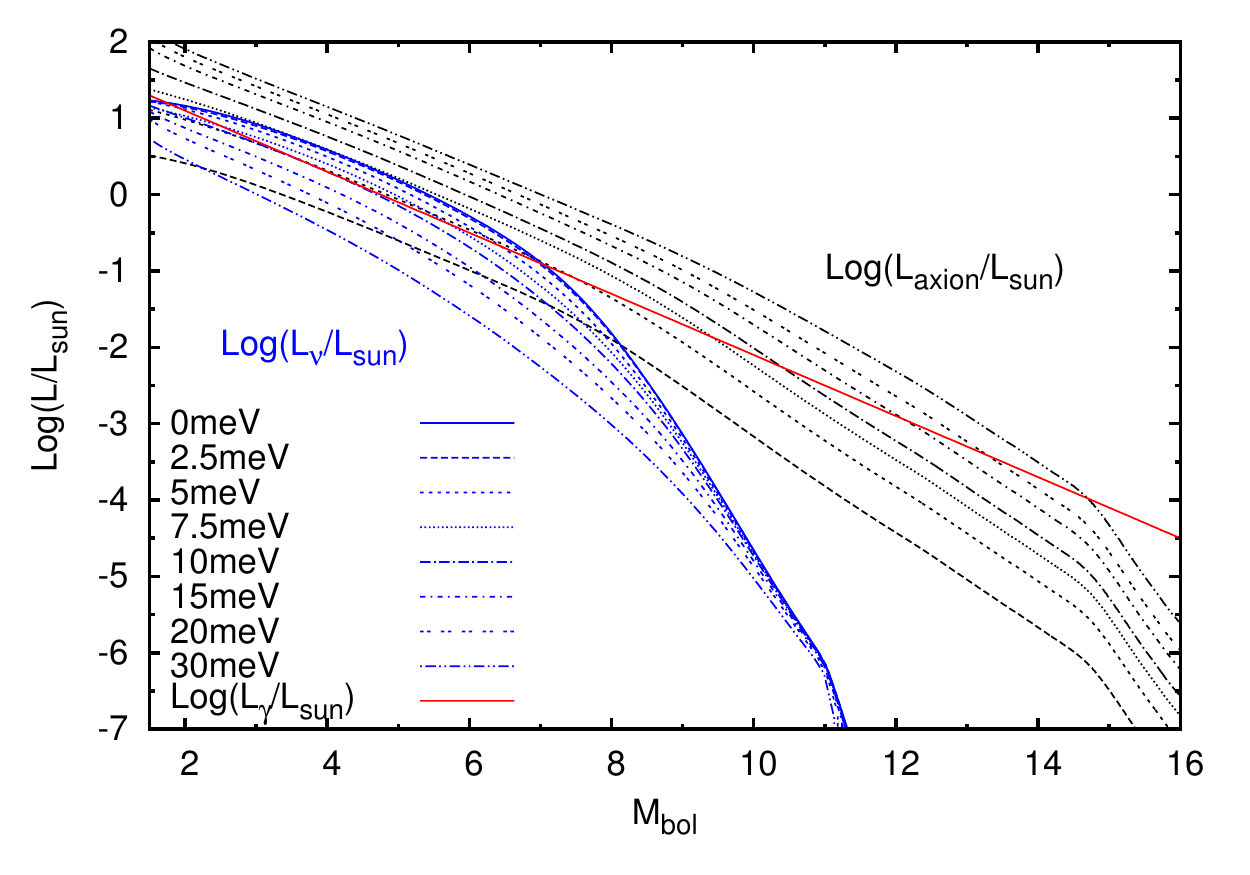} 
\caption{ Axion (black curves) and neutrino (blue curves) emission for our 0.609$M_\odot$
  sequences for different axion masses. The impact of the axion
  emission in the thermal structure of the white dwarf can be
  appreciated in the decrease of the neutrino emission at higher axion
  masses. Clearly, axion emission can not be treated perturbatively at
  $m_a\cos^2\beta>5$meV.}
\label{Fig:Lumi}
\end{figure*}

 Fig. \ref{Fig:Lumi} shows the axion, photon and neutrino emission for a white
 dwarf of 0.609$M_\odot$ under the assumption of different axion masses. It is
 clear from Fig. \ref{Fig:Lumi} that axion emission leads to a decrease of the
 neutrino emission at the same WD luminosity.  One of the results of our
 computations is the realization that the feedback of the axion emission into
 the neutrino emission can not be neglected in the range of axion masses
 suggested by asteroseismological determinations ($m_a>10$ meV,
 \cite{2012MNRAS.424.2792C, 2012JCAP...12..010C}) at relatively high WD
 luminosities ($M_{\rm bol}\lesssim 8$). This result is due to the fact that
 when axion emission is included this leads to an extra cooling of the white
 dwarf core which alters the thermal structure of the white dwarf, as compared
 with the case without axion emission. This, in turn, leads to a decrease of
 the neutrino emission at a given surface luminosity of the star. The main
 consequence of this is that WD cooling is less sensitive to the existence of
 axions, than a perturbative approach would suggest, due to the additional
 energy loss due to the existence of axions is then counterbalanced by the
 decrease of the energy lost by neutrino emission.  Note that for axion masses
 as small as $m_a\cos^2\beta=5$ meV the neutrino emission is already affected
 by the axion emission, being different from the case with no axions
 ($m_a\cos^2\beta=0$ meV).  Consequently the axion emission needs to be
 treated self-consistently for high luminosity WDs ($M_{\rm bol}\lesssim 8$)
 when dealing with axions in the range claimed by asteroseismology or
 detectable through the white dwarf luminosity function. At lower luminosities
 ($M_{\rm bol}\gtrsim 8$) the neutrino emission becomes negligible and thus
 the cooling speed is unaffected by the feedback of axion emission.  Clearly,
 at low luminosities the feedback of axion emission into the total energy loss
 of the star is negligible.

\begin{figure}[]
\includegraphics[clip, angle=0, width=15.cm]{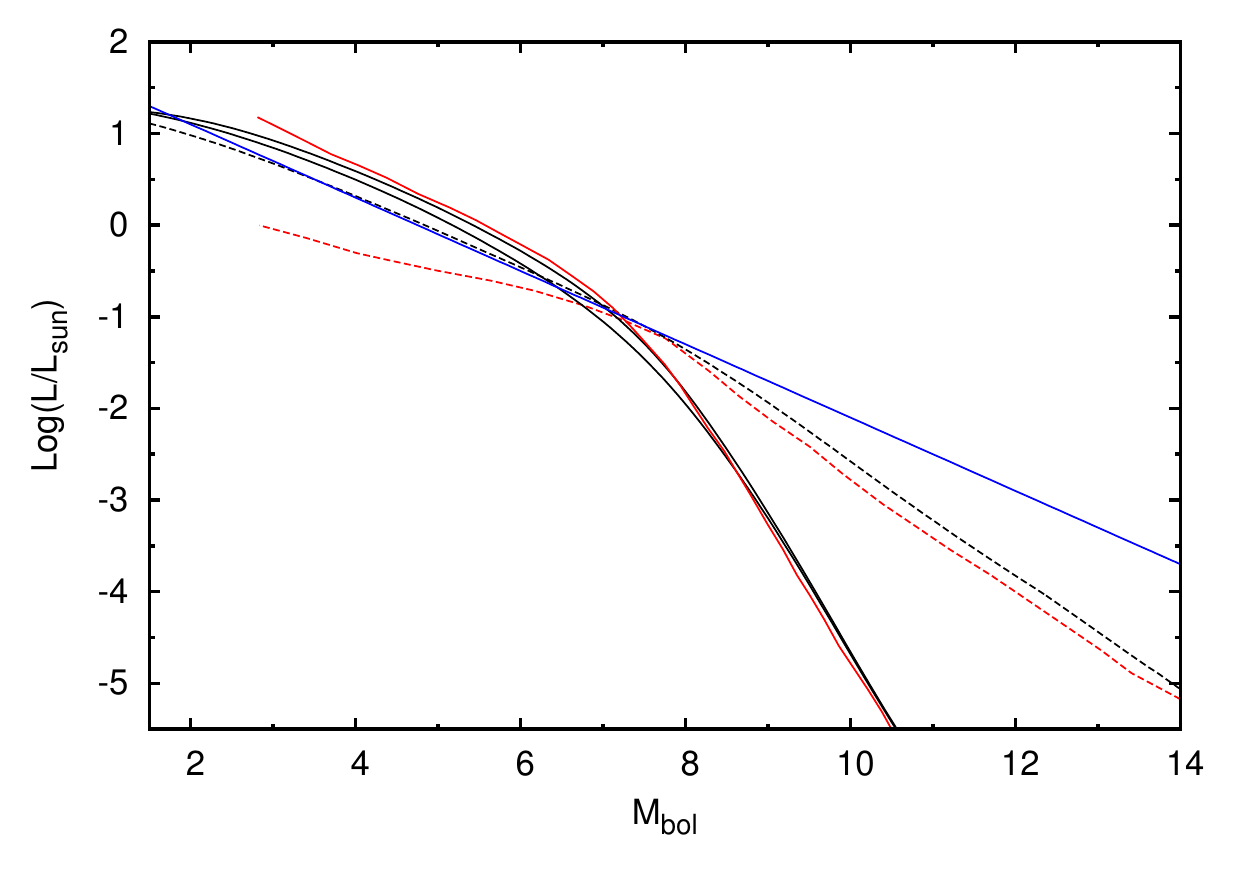} 
\caption{ Comparison of the axion ($m_a\cos^2\beta=5$meV) and neutrino
  emission of our 0.609$M_\odot$ and the 0.61$M_\odot$ sequence of
  \cite{2008ApJ...682L.109I}. The black solid lines represent the neutrino
  luminosity (from top to bottom: $m_a\cos^2\beta=0,\,5\mathrm{meV}$) for our
  model. The red solid line represents the neutrino emission of Isern
  model. The black dashed line represents the $5\mathrm{meV}$ axion emission
  for our model and the red dashed line represents the $5\mathrm{meV}$ axion
  emission of \cite{2008ApJ...682L.109I} model. The blue solid line represents
  the photon luminosity.  The effects of the departure from the isothermal
  core approximation can be appreciated at $M_{\rm bol}<7$.}
\label{Fig:Lumi1}
\end{figure}
In Fig. \ref{Fig:Lumi1} the axion and neutrino emission of our 0.609$M_\odot$
sequence is compared with the 0.61$M_\odot$ sequence of
\cite{2008ApJ...682L.109I} for an axion of $m_a\cos^2\beta=5$ meV. There is an
overall good agreement for axion emission between both predictions at low
luminosities ($M_{\rm bol}\gtrsim 7$). The departure between both curves at
high luminosities ($M_{\rm bol}\lesssim 7$) can be traced back to the
isothermal core approximation of \cite{2008ApJ...682L.109I} which leads to an
underestimation of the axion emission at high luminosities when the maximum
temperature of the core is located off-centered. Also, as mentioned before, to
obtain accurate cooling timescales the feedback of the axion emission into the
thermal structure of the white dwarf is only relevant for high luminosity WDs
($M_{\rm bol}\lesssim 8$) and thus, only marginally relevant for the range of
luminosities studied in most of the present work ($7 \leq M_{\rm Bol}\leq 12.5$).
Consequently, for the aim of comparing theoretical and observed WDLFs in the
range $7 \leq M_{\rm Bol}\leq 12.5$ the impact of the approximations
adopted by \cite{2008MmSAI..79..545I} and \cite{2008ApJ...682L.109I} are not
significant.

\section{Impact of the axion emission in the WDLF} 
\label{axionWDLF}

\begin{figure*}[ht]
\includegraphics[clip, angle=0, width=15cm]{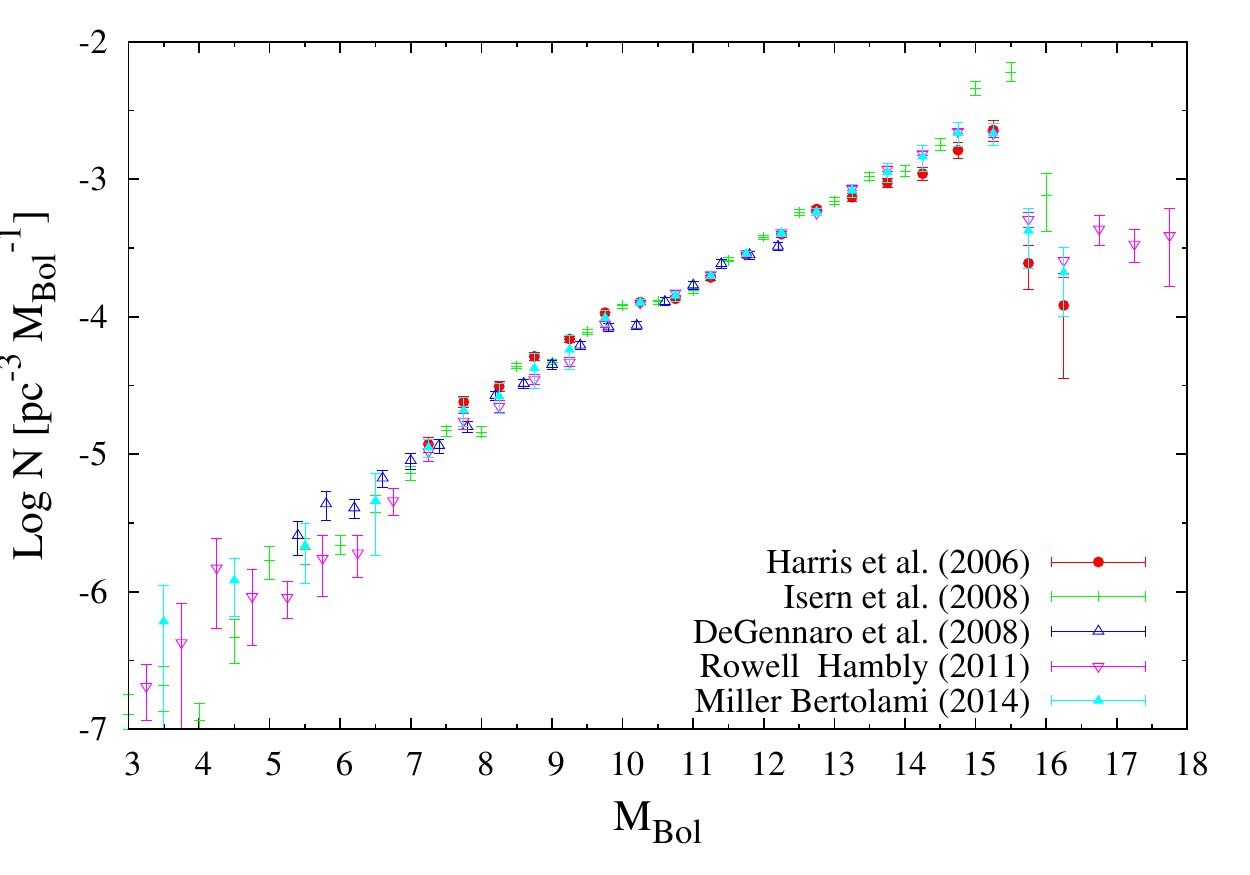} 
\caption{ White Dwarf luminosity functions of the Galactic disk adopted for
  comparison in this work. In this plot the WDLF from
  \cite{2011MNRAS.417...93R} has been multiplied by a factor of 1.862 as
  described in \cite{2014A&A...562A.123M}.}
\label{Fig:WDLF_obs}
\end{figure*}

In order to take into account possible systematics in the comparison and also
to allow for a direct comparison with previous works, we have adopted five
different derivations of the WDLF of the Galatic disk.
Specifically, we compare our theoretical WDLFs with those derived, or
constructed, by
\cite{2006AJ....131..571H, 2008AJ....135....1D, 2008ApJ...682L.109I,
  2011MNRAS.417...93R, 2014A&A...562A.123M}, see
Fig.  \ref{Fig:WDLF_obs}. While  the WDLF of  \cite{2006AJ....131..571H} was derived from
the SDSS-DR3 using the reduced proper motion technique without separeating
them into H-rich and H-deficient white dwarfs, the WDLF of
\cite{2008AJ....135....1D} was derived from the SDSS-DR4 but constraing it to
spectroscopically identified H-rich white dwarfs.  In order to
allow for a direct comparison of our results with those presented by
\cite{2008ApJ...682L.109I}, we have also included in our study the WDLF
presented by \cite{2008ApJ...682L.109I}\footnote{This WDLF was constructed
  from the DR3-SDSS data (Harris, March 2005, private communication) with
  $V_{\rm tan}>20$ km/s.}. All the previous WDLF come from the SDSS survey so
we also included in our analysis the, completely independent, WDLF derived by
\cite{2011MNRAS.417...93R} from the SuperCOSMOS Sky Survey. Finally, to make
full use of the new WDLFs that extend to the high luminosity regime, we have
also included the WDLF constructed by \cite{2014A&A...562A.123M} from two sets
of completely independent WDLFs \citep{2009A&A...508..339K, 2011MNRAS.417...93R}.

\begin{figure}[ht]
\includegraphics[clip, angle=0, width=15cm]{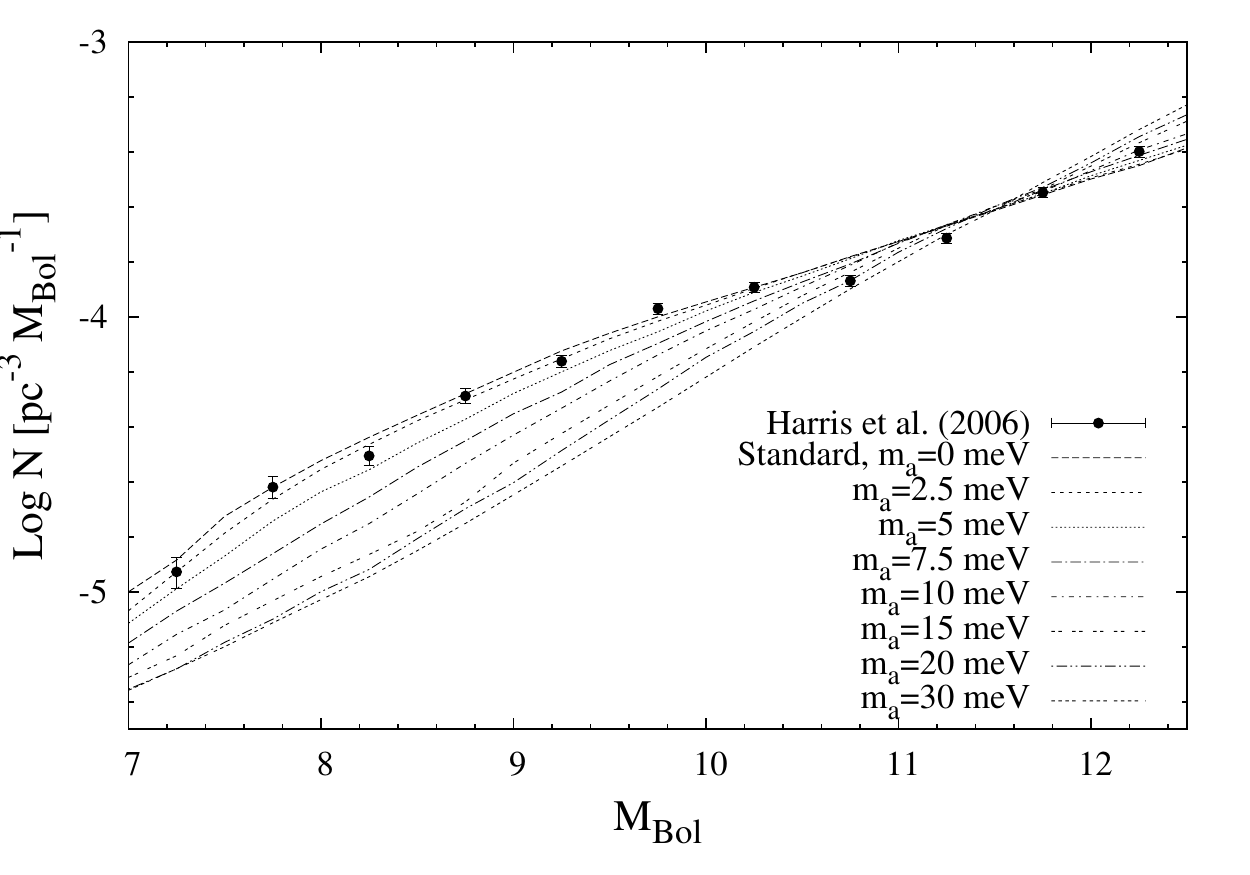} 
\caption{ White Dwarf luminosity function constructed for the
    different axion masses compared with the luminosity function
    derived by \cite{2006AJ....131..571H}. DFSZ-axions heavier than $m_a>10$meV
    are clearly excluded by the observed white dwarf luminosity
    functions. Note that theoretical WDLF constructed with $m_a\ge
    10$meV almost overlap each other in the range of interest. This is due to
    the fact that the normalization region is, then, also affected by
    the axion emission.}
\label{Fig:WDLF}
\end{figure}
 
\subsection{WDLFs at intermediate luminosities}

 In Fig. \ref{Fig:WDLF} we show, as an example, the resulting white dwarf
 luminosity functions for each axion mass as compared with the WDLF derived by
 \cite{2006AJ....131..571H}.  It can be clearly seen that axion masses larger
 than 10 meV would lead to apparent disagreements with the luminosity function
 derived by \cite{2006AJ....131..571H}. Large axion masses are also in
 disagreement with the WDLFs of \cite{2008AJ....135....1D},
 \cite{2008ApJ...682L.109I} and \cite{2011MNRAS.417...93R}. To have a
 quantitative measure of the agreement between the WDLF computed under
 different assumptions and the observational WDLFs (Fig. \ref{Fig:WDLF_obs}),
 we have performed a $\chi^2$-test. The value of $\chi^2$ was computed as
\begin{equation}
\chi^2({m_a})=\frac{1}{(N-1)}\sum_{i=1}^N \frac{(n_i^{\rm O}-n_i^{m_a})^2}{{\sigma_i}^2},
\label{eq:chi}
\end{equation}
where $N=11$\citep{2006AJ....131..571H, 2011MNRAS.417...93R}, 13
\citep{2008AJ....135....1D} or 10 \citep{2008ApJ...682L.109I}, $n_i^{\rm O}$
stands for the values derived in each observational WDLF and $n_i^{m_a}$ is
the theoretically computed number of stars under the assumption of different
axion masses ($m_a$). We have estimated ${\sigma_i}$ as those presented in
each observational WDLF, which means that we are neglecting errors coming from
the uncertainties in the theory of stellar evolution. The $N-1$ in
eq. \ref{eq:chi} takes into account the fact that the $n_i^{m_a}$ values are
not completely independent of the observations, as they are normalized to fit
the observations, as described in eqs. \ref{eq:norm1} and \ref{eq:norm2}.

The $\chi^2-$values obtained from Eq. \ref{eq:chi} are, in all the cases, too
large, according to $\chi^2-$ test, implying that a significant disagreement
exists between all the derived theoretical WDLFs and the observations. The
reasons for this disagreement seems to be two-folded. On the one hand the
different WDLFs are not consistent between themselves, suggesting that the
uncertainties are larger than quoted in the WDLFs. On the other hand, our
method does not quantify the uncertainties in the constructed WDLFs. In
particular, very recent short term fluctuations in the late SFR of the Galaxy
might introduce sizeable departures from the assumption of a constant SFR.
Fortunately, although the derived $\chi^2-$values are large, significant
differences of more than one order of magnitude exist between the best-fit
$\chi^2-$values and the rest. 

\begin{figure*}
\includegraphics[clip, angle=0, width=8cm]{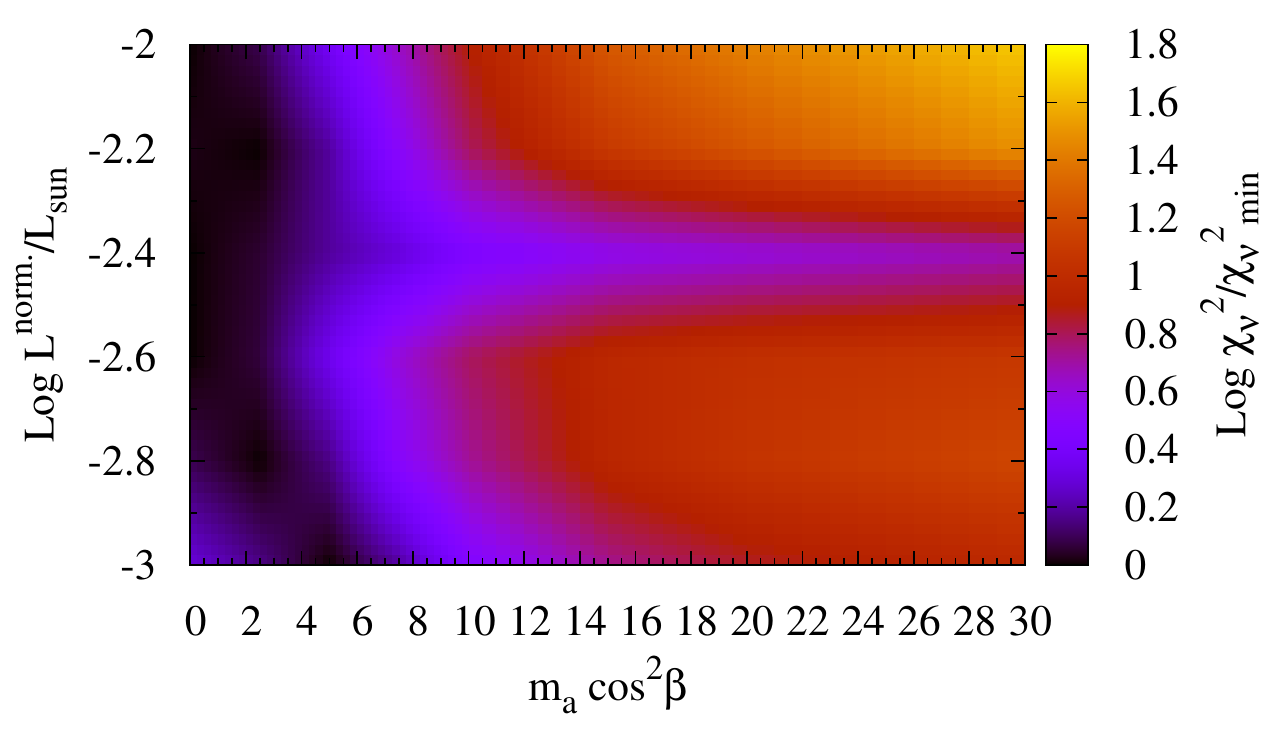}
\includegraphics[clip, angle=0, width=8cm]{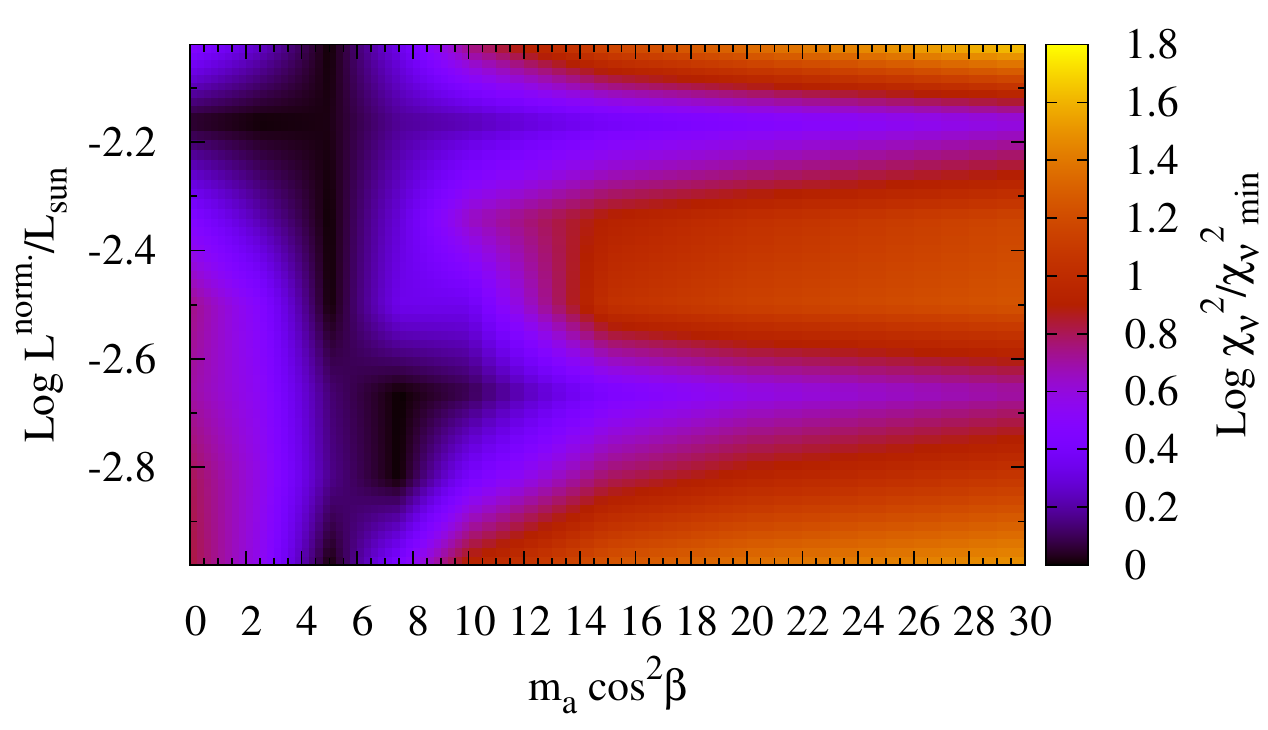}\\
\includegraphics[clip, angle=0, width=8cm]{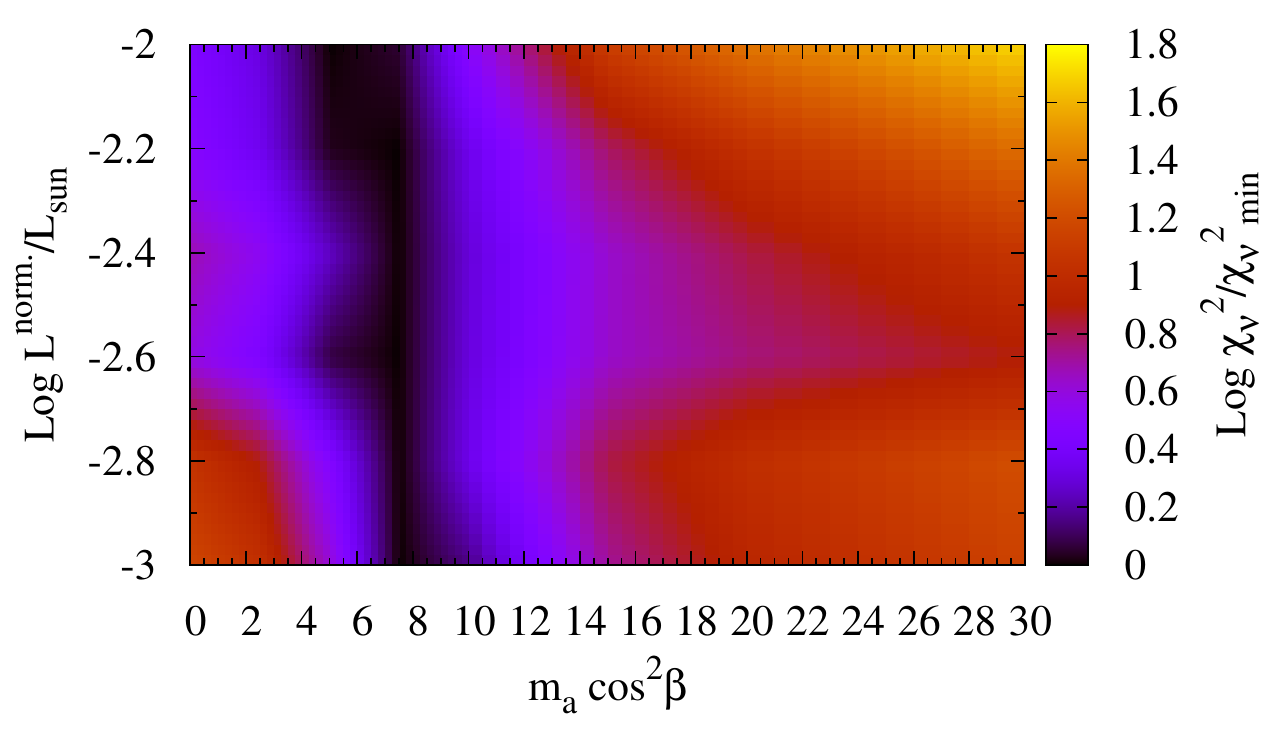}
\includegraphics[clip, angle=0, width=8cm]{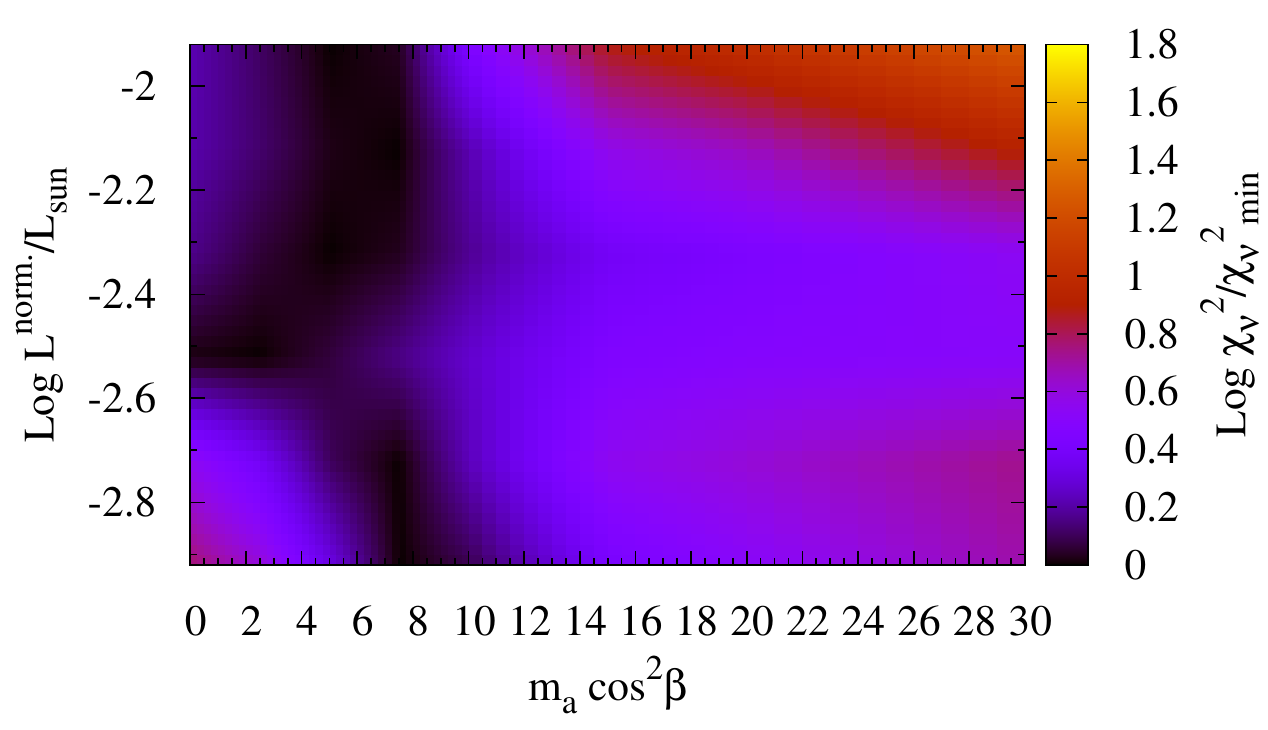}
\caption{$\chi^2$-values (color map) derived from the comparison of the
  theoretical WDLFs with those obtained by \cite{2006AJ....131..571H} (top
  left), \cite{2008AJ....135....1D} (top right), \cite{2011MNRAS.417...93R}
  (bottom left) and \cite{2008ApJ...682L.109I} (bottom right), for different
  choices of the normalization point (y-axis) and the mass of the axion
  (x-axis).}
\label{Fig:chi2}
\end{figure*}

In Fig. \ref{Fig:chi2} we show the $\chi^2-$values relative to the best fit
value for each normalization point and for each WDLF. As apparent from
Fig. \ref{Fig:chi2}, while the WDLF of \cite{2006AJ....131..571H} points
towards values of the axion mass $m_a\cos^2\beta\lesssim 2.5$ meV the WDLFs
from \cite{2008AJ....135....1D} and \cite{2011MNRAS.417...93R} favour the
existence of some extracooling implying axion masses of 5 meV$\lesssim
m_a\cos^2\beta \lesssim 7.5$ meV. It is particulartly worth noting that the
essence of this result is true independent of the normalization point, or
normalization method, of the theoretical WDLF. In fact, when we adopt the
second normalization method for the WDLF (Eq. \ref{eq:norm2}) we reobtain the
same global result.  The different favoured axion mass value should not be surprising, as it is
apparent from Fig. \ref{Fig:WDLF_obs} that the differences between WDLFs are
beyond the quoted error bars. In addition, when we compare with the WDLF
adopted by \cite{2008ApJ...682L.109I} (which is a preliminar version of that
in \cite{2006AJ....131..571H}) we reobtain the main results quoted in that
article, that the best fit models imply values of the axion mass in the range
2.5 meV$\lesssim m_a\cos^2\beta\lesssim 7.5$ meV, depeding on the
normalization point/method. Note, however, that in this case differences in
the $\chi^2-$values are smaller than in the three previous cases and less
stringent constraints can be drawn.

\subsection{Constraints from the WDLF at $3<M_{\rm Bol}<12.5$}
\begin{figure*}
\includegraphics[clip, angle=0, width=15cm]{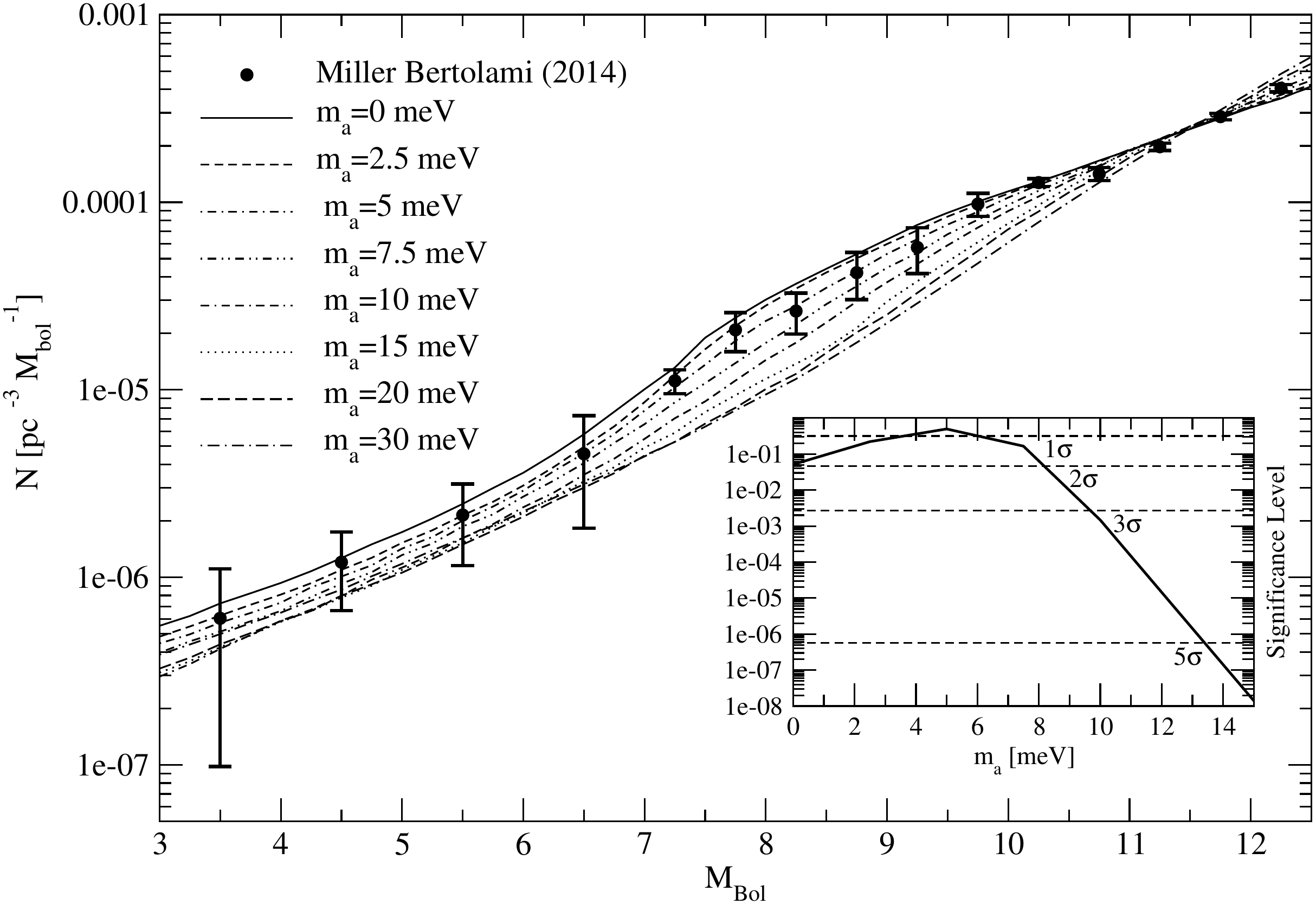}
\caption{Comparison of the WDLF constructed by \cite{2014A&A...562A.123M} with
  the theoretical WDLF for different axion masses. The continuous line in the
  inset shows the significance level at which different values of the axion
  mass are discarded by a $\chi^2$-test. The significance levels associated
  with different $\sigma$-values are show for reference with short dashed
  lines.}
\label{Fig:WDLF-hot}
\end{figure*}

Fig. \ref{Fig:WDLF-hot} shows the comparison between the theoretical WDLF
computed for different axion masses with the WDLF of the Galactic disk
constructed by \cite{2014A&A...562A.123M}. This WDLF was constructed from the
WDLF determined by the SDSS \citep{2006AJ....131..571H,2009A&A...508..339K}
and the SuperCOSMOS \citep{2011MNRAS.417...93R} sky surveys. The size of
the error bars reflects the discrepancies between both WDLFs. Consequently, the
error bars not only reflect internal statistical uncertainties but also
systematic discrepancies between both WDLF ---see \cite{2014A&A...562A.123M}
for a detailed discussion on these issues. In addition, this WDLF includes
information from the WDLF at much higher luminosities where axions are expected
to be an important cooling mechanism even for low axion masses. As in the
previous cases we have compared the theoretical and observational WDLFs by
adopting different normalization points and methods (Eqs. \ref{eq:norm1} and
\ref{eq:norm2}). As it happened in the previous cases (Fig. \ref{Fig:chi2})
for the first normalization method the best fit theoretical luminosity
function depends slightly on the adopted normalization point, but the main
result is stable: A $\chi^2-$test marginally prefers values around
$m_a\cos^2\beta\lesssim 5$ meV while axion masses of $m_a\cos^2\beta\lesssim
10$ meV are excluded at a high significance level. This can be seen in
Fig. \ref{Fig:WDLF-hot} where the results in the case of the second
normalization method are shown. As can be seen in the inset, while the best
fit model corresponds to $m_a\cos^2\beta\lesssim 5$ meV all models below
$m_a\cos^2\beta\lesssim 7.5$ meV can not be rejected at a $2\sigma$-like
confindence level. On the contrary, all theoretica WDLF constructed with
$m_a\cos^2\beta\gtrsim 10$ meV are strongly at variance with the WDLF
from \cite{2014A&A...562A.123M}.

\section{Discussion and conclusions}
\label{conclusion}
 In order to place constraints on the possible coupling strength between
 axions and electrons ($g_{ae}$) we have performed a detailed study of the
 impact of the axion emission on the cooling of white dwarfs.  In particular,
 we improved previous works by including a self consistent treatment of the
 axion emission into the thermal structure of the white dwarf models, and by
 including several different WDLFs in the analysis. In particular, for the
 first time we extended the comparison to the high luminosity regime where
 axions are expected to be the main cooling channel.  This was performed on
 the basis of state of the art initial white dwarf models and microphysics.
 In addition, we tested the dependence of the comparison on the choice of the
 normalization points of the theoretical WDLFs, as well as on the
 normalization method adopted. We quantitatively weighted the agreement
 between theory and observations by means of a $\chi^2$-fit.

The main results of the present work can be summarized as follows. In the
luminosity range $7 \leq M_{\rm Bol}\leq 12.5$ we find an overall good
agreement between the perturbative approach adopted by
\cite{2008ApJ...682L.109I} and the result of the self-consistent
full-evolutionary computations presented here. On the contrary, for the range
$M_{\rm Bol}\leq 7$ both the feedback of the axion emission on the thermal
structure of the white dwarf and the departure from the isothermal core
approximation need to be taken into account. This is worth noting in the light
of new determinations of the WDLF which extend to higher luminosities and
temperatures.  From an inspection of the results in Figs. \ref{Fig:chi2} and
\ref{Fig:WDLF-hot} we can conclude that the values of $m_a\cos^2\beta\simeq
17$ meV inferred from the asteroseismology of G117-B15A and R548
\citep{2012MNRAS.424.2792C, 2012JCAP...12..010C} are significantly disfavoured
by the study of the WDLF.  It is worth noting that the high mass of the axion
derived in those works is a direct consequence of the identification of the
215 s (213 s) mode of G117-B15A (R548) as a mode trapped in the
envelope. Consequently, our result can also be viewed as a strong argument
that those modes are not trapped modes. This is true independent of the
normalization of the theoretical WDLF. When comparing with the WDLF of
\cite{2014A&A...562A.123M} that includes information from an extended range of
luminosities ($3<M_{\rm Bol}<12.5$) we find that values of
$m_a\cos^2\beta\gtrsim 10$ meV are strongly at variance with the WDLF. In
particular a $\chi^2-$test suggests that values of $m_a\cos^2\beta\gtrsim 10$
meV are discarded at a $3\sigma$ confidence level. Unfortunately, it is not
possible to attach a credible confidence level to all these results, as
discrepancies between different observed WDLF, as well as a lack of a proper
estimation of the uncertainties in the SFR, prevent us from doing so.  It
should be noted, however, that uncertainties in the theoretical WDLFs might
not be dominant. In fact, numerical experiments suggest that possible
fluctuations of the order of 50\% in the SFR during the last 1.5 Gyr, as those
suggested by \cite{2013MNRAS.434.1549R}, would only translate into an
uncertainty of less than a 10\% in the WDLF. In addition, current
uncertainties in the cooling times of white dwarfs are below a 10\%
\citep{2010ApJ...716.1241S, 2013Salaris} and would imply a similar uncertainty
in the WDLF. Uncertainties of the order of 10\% in the theoretical WDLF are
not expected to affect significantly the constraints derived in Sect. 4.2. On
the other hand discrepancies between different observed WDLF are
significant. Note, for example that, while the WDLF of
\cite{2006AJ....131..571H} favours almost no extra cooling mechanism
($m_a\cos^2\beta\lesssim 2.5$ meV) axion masses below $m_a\cos^2\beta\simeq
10$ meV are not excluded by our current knowledge of the WDLF of the Galactic
disk. In particular, it is worth noting that the features in some WDLFs
\citep{2008AJ....135....1D, 2011MNRAS.417...93R} can be interpreted as
suggestions for axion masses in the range 2.5 meV$\lesssim
m_a\cos^2\beta\lesssim 7.5$ meV. This is particularly interesting in the light
of the forthcoming International Axion Observatory
(IAXO,\cite{2013arXiv1302.3273V}) which will be able to explore axion masses
in the range $m_a\cos^2\beta\gtrsim 3$ meV.

 Needles to say, more work needs to be done. In particular
completely independent WDLFs, like those presented by
  \cite{2009A&A...508..339K, 2011MNRAS.417...93R, 2013ASPC..469...83K}, are
  needed in order to account for possible systematics in the
  determinations. In particular, the new WDLFs being derived from the
  DR10-SDSS (Kepler, private communication), which will include a much larger
  range of WD luminosities will significantly improve our constraints. On the
  theoretical side a more complex approach
  (e.g. \cite{2006MNRAS.369.1654G}) is needed to account for the
  theoretical uncertainties in the construction of the WDLF. In particular
  the impact of the uncertainties in the SFR, Galactic dynamics and the
  H-rich/H-deficient white dwarf ratio need to be assessed.

\bibliographystyle{JHEP}

\providecommand{\href}[2]{#2}\begingroup\raggedright\endgroup

\acknowledgments
 The authors thank H. Harris, S. DeGennaro, N. Rowell and
  N. Hambly for the data of their respective WDLF. We also thank G. Raffelt
  and A. C\'orsico for reading a preliminar version of the manuscript and
  making useful suggestions. This research was supported by PIP
  112-200801-00940 from CONICET and through the Programa de Modernizaci\'on
  Tecnol\'ogica of the ANPCYT, BID 1728/OC-AR. M3B is supported by a
  fellowship for postdoctoral researchers from the Alexander von Humboldt
  Foundation.


\end{document}